\documentclass[12pt,preprint]{aastex}

\usepackage{epsfig,amsmath,subfigure}

\def\gsim{\;\rlap{\lower 2.5pt
 \hbox{$\sim$}}\raise 1.5pt\hbox{$>$}\;}
\def\lsim{\;\rlap{\lower 2.5pt
   \hbox{$\sim$}}\raise 1.5pt\hbox{$<$}\;}
\newcommand{\be}{\begin{equation}}
\newcommand{\beq}{\begin{equation}}
\newcommand{\ba}{\begin{eqnarray}}
\newcommand{\ee}{\end{equation}}
\newcommand{\eeq}{\end{equation}}
\newcommand{\ea}{\end{eqnarray}}

\shortauthors{Bower et al.}
\shorttitle{Tidal Disruptions}

\begin{document}

\title{Constraining the Rate of Relativistic Jets from Tidal Disruptions Using Radio Surveys}

\author{Geoffrey C. Bower\altaffilmark{1}}
\altaffiltext{1}{University of California, Berkeley, Radio Astronomy Laboratory and Department
of Astronomy, 601 Campbell Hall \#3411, Berkeley, CA 94720, USA; gbower@astro.berkeley.edu }

\begin{abstract}
Tidal disruption of stars by massive black holes produce transient
accretion flows that flare at optical, UV, and X-ray wavelengths.  At late times, these
accretion flows may launch relativistic jets that can be detected 
through the interaction of the jet with the dense interstellar medium
of the galaxy.  We present an upper limit
for the flux density of a radio counterpart to a tidal disruption event 
detected by GALEX that is a factor of 6 below theoretical predictions.
We also examine existing radio surveys for 
transients with a time scale of 1 year and use these to set a $2 \sigma$ 
upper limit
on the rate of tidal disruption events producing 
relativistic jets of $\sim 14 \times 10^{-7} {\rm\ Mpc^{-3}\ y^{-1}}$.
This rate is an order of magnitude lower than the highest values
from theoretical models and is consistent with detection rates
from optical and X-ray surveys.
\end{abstract}
\keywords{radio continuum:  general --- galaxies:  active ---
radio continuum:  galaxies --- surveys}
%% VLA code:  AF420

\section{Introduction}

Massive black holes at the centers of galaxies disrupt stars
that pass within the tidal radius of the black hole, resulting 
in a process of transient accretion
\citep[e.g.,][]{1988Natur.333..523R}.  Dynamical models predict that
these processes occur every $10^3$ to $10^5$ y per galaxy
\citep{2004ApJ...600..149W}.  The
accretion process leads to a flare that peaks at X-ray or UV 
wavelengths and has a power-law decline with a characteristic timescale
of months.  Such events are recognized by the flare 
and the transition from a normal galaxy to a state with an AGN spectrum.
\citet{2009ApJ...698.1367G} summarize the current state of our knowledge
regarding tidal disruption events detected at X-ray, UV, and optical
wavelengths.  In total, 10 events have been characterized.  Synoptic
optical surveys may detect tens of such events and prove 
powerful in characterizing tidal disruption events
and determining their statistics \citep{2010arXiv1008.4131S,2010arXiv1009.1627V}.

A significant astrophysical problem is the coupling between
accretion and the launching of jets.  Such systems are ubiquitous
yet there is no complete theory that can predict jet properties
based on a known accretion flow.  Tidal disruption events represent
a unique laboratory for exploring the accretion-outflow coupling
under a range of different accretion conditions over a short period of time.  
If accretion leads to a relativistic outflow, this will likely radiate
synchrotron emission, which can be detected at radio wavelengths.
A recent model explores radio emission that results from the interaction
of the outflow with the interstellar medium 
\citep{2011arXiv1102.1429G}.  
Radio surveys have the advantage that they are unaffected by extinction
in the dense nuclear environments of the host galaxy;
however, there are no known 
detections of radio emission from tidal disruption events.

In this paper we examine what constraints that existing and
future radio surveys can place on tidal disruption events.  In \S~\ref{sec:theory},
we briefly summarize a new model for radio emission from tidal disruptions.  
In \S~\ref{sec:surveys}, we discuss the relevant survey data and identify
candidate events.  In \S~\ref{sec:results}, we use these results to set 
constraints on the rate of tidal disruptions that produce relativistic
jets.  In \S~\ref{sec:galex}, we present
archival Very Large Array results for a tidal disruption event detected by
GALEX.  In \S~\ref{sec:summary}, we summarize our results.

\section{Radio Emission from Jet-Induced Reverse Shocks \label{sec:theory}}

\citet[][hereafter, GM]{2011arXiv1102.1429G}
introduce a model in which a reverse shock from the interaction of a
jet with the dense interstellar medium of the galaxy produces an afterglow.
The afterglow has a peak flux density
\begin{equation}
F=2 C  \left( { D \over 1 {\rm\ Gpc} } \right)^{-2} {\rm\ mJy},
\end{equation}
where $D$ is the distance to the source and $C$ is a constant of
order unity that depends on details of the jet and ISM physics.
The spectrum peaks at frequency of $\sim 25$ GHz and extends
as $\nu^{1/3}$ to lower frequencies before reaching the self-absorption
frequency.  As the source cools,
the peak flux density propagates to lower frequencies with proportionality $\nu^{1/2}$.
The reverse shock spectrum is expected to peak $\sim 1$ year
following the tidal disruption event and have a duration of 1 year or less.
This afterglow has the important property that it emits isotropically.
Accordingly, events are not limited by the beaming angle of the relativistic
jet that creates the relativistic shock.

A significant challenge in identification of a tidal disruption event
in a nearby galaxy may be confusion with galactic emission.  In this
case, the tidal disruption would not appear as a true transient but rather
as significant variability in a detected source.  The amplitude of this
variability must be large in the model of GM, however.  For example,
the galaxy NGC 891 has a flux density of 700 mJy at 1.4 GHz.  If moved 
from its current
distance of 7 Mpc to 1 Gpc, it would have a flux density of 35 microJy,
a factor of 10 below the predictions of GM for a tidal flare at this distance.  
Thus, in practice
tidal disruption events will appear as transients or extreme
variables in most surveys.

\section{Survey Results \label{sec:surveys}}

We summarize here existing radio surveys that have sensitivity on timescales 
of $\sim 1$ year and are, therefore, sensitive to tidal disruption events.
In Table~\ref{tab:surveys}, we give the frequency ($f$), limiting flux density
($F_{lim}$),
survey area ($\Omega$), number of epochs ($N_e$), and number of
candidate events ($N_{can}$).  The number of epochs is 
the number of independent year-long observations.  

The archival VLA survey \citep{2007ApJ...666..346B} at 5.0 and 8.4 GHz observed a single pointing
weekly for 20 years, producing nearly 1000 independent epochs.  
\citet{2007ApJ...666..346B} produced maps and searched for transients on
time scales of a single epoch, 2 month integrations, and 1 year integrations.
No radio transients were detected from the 1-year integrations.
Two radio transients were detected from the 2-month integrations but neither
is associated with the nucleus of an optically-detected galaxy.
A characteristic sensitivity of 0.1 mJy is chosen to represent the overall
survey at 2-month and 1-year sensitivity.  At this flux density, 
the effective survey area over all epochs
is 0.26 deg$^2$ at 1 year.  This quantity is equal to $N_e \Omega$.  Dividing over $N_e=20$,
we find the per epoch area of 0.013 deg$^2$.  Limits for two-month integrations are
similar to the 1-year results.
Angular resolution of the survey
ranged from $<1$\arcsec to $\sim 15$\arcsec.

An analysis of 3C 286 observations from the VLA archives at 1.4 GHz was
presented by \citet{2011ApJ...728L..14B}.  This paper explored daily variations in data obtained
over 23 years in 1852 epochs.  Since 
integrations longer than a single day were not produced,
the single epoch flux density threshold sets the transient sensitivity limit.
This limit is relatively poor, limiting the overall sensitivity of this result
to month- or year-long events.  These data were taken in the C and D configurations of
the VLA, giving characteristic angular resolution $>15$\arcsec.  No radio
transients were found above a characteristic flux density of 70 mJy.
The solid angle is set by the field of view imaged.

The PiGSS-I \citep{2010ApJ...725.1792B} and ATATS-I \citep{2010ApJ...725.1792B} results are from Allen Telescope
Array \citep{2009IEEEP..97.1438W} single-epoch surveys at 3.1 and 1.4 GHz, respectively.  These
survey results are compared against the NVSS catalog 
\citep{1998AJ....115.1693C} to determine the
rate of transient sources.  The characteristic timescale of both of these
efforts is any timescale longer than the integration timescale of NVSS, which is
less than 1 minute.  NVSS observations were conducted more than 1 decade earlier
than the ATA observations.  The PiGSS-I catalog is constructed from observations
spread over 4 months in 2009 and has a resolution of 100\arcsec.  The ATATS-I
catalog is constructed from observations spread over 4 months in 2008 and has 
a resolution of $\sim 200$\arcsec.

Light curves of sources detected in the VLA, 3C 286, PiGSS-I, and ATATS-I
surveys indicate no variability greater than a factor of a few in all surveys
on long timescales.

\citet{2002ApJ...576..923L} and \citet{2006ApJ...639..331G} presented analyses of a comparison of the VLA FIRST
\citep{1995ApJ...450..559B}
and NVSS 1.4-GHz catalogs in search of radio transients.  The characteristic separation
time between these catalogs is $\sim 1$ year.  A radio supernova was the only
transient identified.  The characteristic resolution
of the two surveys was 5 and 45\arcsec, respectively.  The authors report
no evidence of extreme variability in sources in either epoch.

\citet{2010arXiv1011.0003B} searched the archives of the Molongolo Synthesis Telescope to
identify all fields observed multiple times over a 22-year period.
The total surface area surveyed was 2776 deg$^2$ at a flux density
threshold of 6 mJy.  All MOST observations
were conducted at 0.843 GHz with a resolution of 45\arcsec.  
The heterogeneous nature of the sample makes
determination of the sensitivity of the survey
at year and longer timescales difficult.
Examination of Figure~5 in their paper indicates that approximately two-thirds
of the observations were paired with others on timescales greater than 1 year.
Accordingly, we scale the total area by a factor of two-thirds.

The MOST survey identified 53 variable and 15 transient sources.  
Of the 15 transient sources from MOST, only three are
unambiguously identified with known objects (SN 1987A, Nova Muscae 1991,
and GRO 1655-40).  Another source is offset from the nucleus of a nearby spiral galaxy
and, therefore, unlikely to be a tidal disruption event.
We can also exclude three sources showing variability on timescales much less
than 1 year.  We further exclude four sources that exhibit multiple 
detections with intervening non-detections.  Two of the remaining four sources 
have optical detections in archival data indicative of a quasar.  The dates
of these optical observations are unclear but the surveys 
from which the data are drawn began and ended several years
prior to the transient detection epoch for both sources.  The authors
argue that these are scintillating AGN.  The remaining two sources 
(J064149-371706 and SUMSS J102641-333615) have a single detection with MOST,
a non-detection with MOST about a year after the original detection,
and no identified optical counterpart.  We note that both sources have apparent
NVSS counterpart that the authors did not identify, indicating that these are
variable sources rather than true transients.  Thus, none of the MOST transients
qualify as plausible tidal disruption candidates.

Additionally, two variable sources (J201524-395949 and J200936-554236)
from MOST are possibly associated with 
the nuclei of spiral galaxies at redshifts of 0.02 and 0.032, respectively.  
J201524-395949
exhibits an increasing flux density over 10 years with amplitude of $\sim 20$ mJy
and a factor of 4 total increase.
J200936-554236 exhibits a 5 mJy rise in flux density over three years (corresponding
to less than factor of 2 increase in flux density).
The amplitude of variability in both sources falls well below the predictions of
GM ($\sim 200$ mJy at this distance).  Nevertheless, these sources
broadly follow the expectations for tidal disruptions and are plausible
candidates.

\section{Limit on the Tidal Disruption Rate Producing Jets\label{sec:results}}

We also tabulate calculated values for 
the limiting distance ($D_{lim}$) at which tidal disruptions would be visible 
under the model 
of GM and an upper limit to the tidal disruption rate producing jets
($r_{TD}$).   $D_{lim}$ is computed
directly from the limiting sensitivity of the surveys and explicitly making use 
of the decreasing peak flux density at lower frequencies ($\propto \nu^{1/2}$).  
We compute the rate from 
\begin{equation}
r_{TD} = { N_{can} \over V N_e},
\end{equation}
where $V$ is the volume surveyed.  
For the case of no candidates, we set the upper limit at $2\sigma$ with $N_{can}=3$.
We give $r_{TD}$ in units of $10^{-7} {\rm\ y^{-1}\ Mpc^{-3}}$.

These estimates clearly indicate that three surveys set the strongest limits.
The VLA archival survey sets limits of $r_{TD} < 29$ for both
two-month and 1-year integrations, respectively.  The FIRST-NVSS
comparison sets an upper limit $r_{TD} < 14$ while the MOST survey gives
a rate limit $r_{TD} \lsim 17$.  Note that in the case of MOST, the
existence of plausible but not convincing
candidates still leaves this value as an upper limit.
Summing over all independent surveys,
we estimate $r_{TD} \approx 4.3$.  We can place an upper limit of $r_{TD} < 14$
at $2\sigma$ confidence.

The radio upper limit is consistent with estimates based on
ROSAT \citep{2002AJ....124.1308D} and SDSS observations
\citep{2010arXiv1009.1627V}.  The latter estimates a rate
$\dot{N} \approx 3 \times 10^{-5} {\rm\ y^{-1}\ galaxy^{-1}}$.
For a space density of galaxies with massive
black holes of $\sim 3 \times 10^{-3} {\rm\ Mpc^{-3}}$, the optical rate
implies $r_{TD} \sim 1.0$.
The limit that we have set falls more than 
an order of magnitude below the most optimistic model
for the event rate of tidal disruptions, which
estimates $r_{TD} \sim 10^{-5} {\rm\ y^{-1}\ Mpc^{-3}}$  \citep{2004ApJ...600..149W}.

Tidal disruption rates for radio detected events are likely to be lower than 
in the optical given that only a fraction of accreting systems typically 
produce radio jets.  Thus, the radio limit can be considered consistent with
the most optimistic optical constraints if $< 10$\% of the systems produce
relativistic jets.  
Similarly, changes in the jet physics model of GM can strongly influence the 
event rate limit.  For instance, the rate $r_{TD}$ scales as the jet luminosity to the -3/2
power.

Future surveys will be capable of significantly improving sensitivity.  As
GM note, the complete PiGSS results have the capability to detect tens
of events.  Non-detections would place significant constraints on $r_{TD}$ and
the reverse shock model.  Similarly, further analysis of data from the VLA
archives can provide significantly more sensitivity.  Year- or month-long integrations of
the 3C 286 data would likely set sensitivity levels at 1 mJy or better while
covering comparable area.  The resulting sensitivity to $r_{TD}$ 
would be in the range of $10$.  Use of additional sources or data at other
frequencies could lead to an order of magnitude improvement.
The VAST intermediate depth survey with ASKAP will achieve 250 microJansky sensitivity over 
$10^4$ deg$^2$ in multiple passes and have substantial sensitivity to phenomena of this kind
\citep{2010AAS...21547012C}.

\section{VLA Images in the Field of GALEX J141929+525206 \label{sec:galex}}

\citet{2006ApJ...653L..25G} reported the detection of a UV flare from an early-type galaxy
with no prior evidence for AGN activity
at z=0.37.  The source was undetected in 2003 June 21 - 29, first
appeared in 2004 March 25 - June 24, and then decayed steadily 
until it was not detected in 2006 March 5 - 7.  

\begin{figure}
\psfig{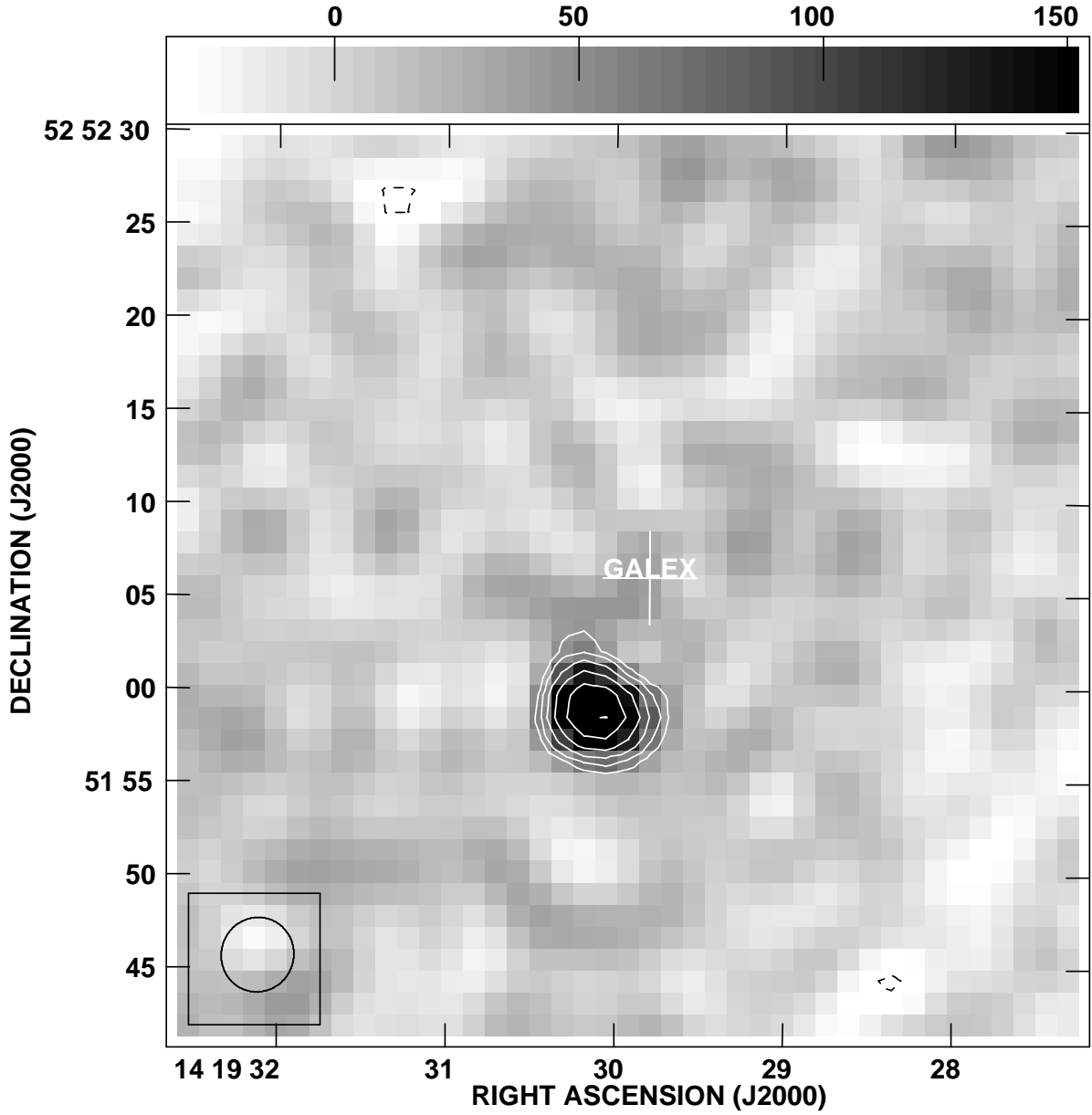}
\caption{Image of the field of the tidal disruption event GALEX J141929+525206. 
The cross marks the position of the GALEX source.  Contours are -3, 3, 4.2, 6, 8.4,
12, 16.8, and 24 times the rms noise of 15 $\mu$Jy.  The grey scale 
ranges from -30 $\mu$Jy to 150 $\mu$Jy. The synthesized beam is
shown in the lower left. 
\label{fig:galex}}
\end{figure}

We used archival VLA data of observations in the field of the tidal flare 
candidate source.  Data were obtained at a frequency of 1.4 GHz 
in spectral line mode to enable full field of view imaging on 
2003 December 2, 2005 April 18, and 2005 June 20 as part of a campaign to
survey the extended Groth strip \citep{2007ApJ...660L..77I}.  
No source was detected in 
these individual epochs at the location of the GALEX source
with image rms of $< 50\ \mu$Jy.
We integrate together all data to produce 
an image with 15 $\mu$Jy rms and 
a synthesized beam $4.0\arcsec \times 3.9\arcsec$  (Figure~\ref{fig:galex}).  
Again, no source is detected in the image of the GALEX source.
A bright, extended source
in the field is associated with the galaxy SDSS J141930.12+525159.

The model of GM predicts a radio flux of 0.2 mJy for this source at a frequency
of 1.4 GHz.  
This is a factor of 6 higher than our $2\sigma$ upper limits in the integrated image.
Thus, our observations, which are coincident with the first appearance of
the UV flare and extend to 1.5 y after the peak, demonstrate that 
this source likely did not produce a relativistic jet that
led to a reverse shock in the interstellar medium, or that the reverse shock
emission was relatively short-lived.

In addition,
we note that \citet{2010arXiv1009.1627V} obtained upper limits for a radio
counterpart of $\sim 0.1$ mJy at 8.5 GHz to tidal disruption event TDE2 in a 
galaxy at a distance of 1.05 Gpc.  
The flux densities fall well below the GM peak-flux predictions of $\sim 1.1$ mJy.
These limits were obtained 7 and 92 days following the first initial optical
detection, and so may not probe the late-time shock event.

\section{Summary \label{sec:summary}}

We have presented here a summary of existing radio surveys with sensitivity
to transients with timescales of $\sim 1$ y.  These surveys provide
the first constraint based on radio data of 
the rate of tidal disruption events around massive
black holes in the nearby Universe.  This upper limit is consistent with
rates set via optical methods.  Note that the rate we have set depends
on details of the GM model for the jet and its interaction with the interstellar
medium of the galaxy.    Factors such as the fraction of systems that 
produce jets, the efficiency with which accretion power is converted to
jet luminosity, and the density of the interstellar medium can significantly
affect the estimated rates.
The non-detection of a radio counterpart to a tidal disruption event 
demonstrates the degree to which we do not understand the accretion-outflow
characteristics of systems of this kind.  

Ongoing surveys such as PiGSS, future surveys such as VAST, and analysis
of archival data can contribute to our constraints on radio properties of
tidal disruptions.  Equally important is follow-up of candidate events
with sensitive observations using the EVLA.  Unambiguous detection 
of a single radio counterpart would provide significant insight.  
Future optical surveys such as those carried out by Pan-STARRS are
predicted to detect $O(10)$ events per year \citep{2009ApJ...698.1367G}.  
Following up all events at a sensitivity 10 times deeper than
the predictions of GM would provide a window into what mechanism in
the accretion-outflow coupling leads to the paucity of radio emission
from tidal disruption events.

\acknowledgements{The National Radio Astronomy Observatory is a facility of the National Science Foundation operated under cooperative agreement by Associated Universities, Inc. This research is supported in part by NSF grant AST-0909245.  The author thanks Joshua Bloom, Eliot Quataert, and Linda
Strubbe for useful discussions.}

%\bibliographystyle{apj}
%\bibliography{myrefs}

\begin{thebibliography}{19}
\expandafter\ifx\csname natexlab\endcsname\relax\def\natexlab#1{#1}\fi

\bibitem[{{Bannister} {et~al.}(2010){Bannister}, {Murphy}, {Gaensler},
  {Hunstead}, \& {Chatterjee}}]{2010arXiv1011.0003B}
{Bannister}, K., {Murphy}, T., {Gaensler}, B.~M., {Hunstead}, R., \&
  {Chatterjee}, S. 2010, ArXiv e-prints

\bibitem[{{Becker} {et~al.}(1995){Becker}, {White}, \&
  {Helfand}}]{1995ApJ...450..559B}
{Becker}, R.~H., {White}, R.~L., \& {Helfand}, D.~J. 1995, \apj, 450, 559

\bibitem[{{Bower} \& {Saul}(2011)}]{2011ApJ...728L..14B}
{Bower}, G.~C., \& {Saul}, D. 2011, \apjl, 728, L14+

\bibitem[{{Bower} {et~al.}(2007){Bower}, {Saul}, {Bloom}, {Bolatto},
  {Filippenko}, {Foley}, \& {Perley}}]{2007ApJ...666..346B}
{Bower}, G.~C., {Saul}, D., {Bloom}, J.~S., {Bolatto}, A., {Filippenko}, A.~V.,
  {Foley}, R.~J., \& {Perley}, D. 2007, \apj, 666, 346

\bibitem[{{Bower} {et~al.}(2010){Bower}, {Croft}, {Keating}, {Whysong},
  {Ackermann}, {Atkinson}, {Backer}, {Backus}, {Barott}, {Bauermeister},
  {Blitz}, {Bock}, {Bradford}, {Cheng}, {Cork}, {Davis}, {DeBoer}, {Dexter},
  {Dreher}, {Engargiola}, {Fields}, {Fleming}, {Forster}, {Gutierrez-Kraybill},
  {Harp}, {Heiles}, {Helfer}, {Hull}, {Jordan}, {Jorgensen}, {Kilsdonk}, {Law},
  {van Leeuwen}, {Lugten}, {MacMahon}, {McMahon}, {Milgrome}, {Pierson},
  {Randall}, {Ross}, {Shostak}, {Siemion}, {Smolek}, {Tarter}, {Thornton},
  {Urry}, {Vitouchkine}, {Wadefalk}, {Weinreb}, {Welch}, {Werthimer},
  {Williams}, \& {Wright}}]{2010ApJ...725.1792B}
{Bower}, G.~C., {et~al.} 2010, \apj, 725, 1792

\bibitem[{{Chatterjee} {et~al.}(2010){Chatterjee}, {Murphy}, \& {VAST
  Collaboration}}]{2010AAS...21547012C}
{Chatterjee}, S., {Murphy}, T., \& {VAST Collaboration}. 2010, in Bulletin of
  the American Astronomical Society, Vol.~42, American Astronomical Society
  Meeting Abstracts \#215, 470.12--+

\bibitem[{{Condon} {et~al.}(1998){Condon}, {Cotton}, {Greisen}, {Yin},
  {Perley}, {Taylor}, \& {Broderick}}]{1998AJ....115.1693C}
{Condon}, J.~J., {Cotton}, W.~D., {Greisen}, E.~W., {Yin}, Q.~F., {Perley},
  R.~A., {Taylor}, G.~B., \& {Broderick}, J.~J. 1998, \aj, 115, 1693

\bibitem[{{Donley} {et~al.}(2002){Donley}, {Brandt}, {Eracleous}, \&
  {Boller}}]{2002AJ....124.1308D}
{Donley}, J.~L., {Brandt}, W.~N., {Eracleous}, M., \& {Boller}, T. 2002, \aj,
  124, 1308

\bibitem[{{Gal-Yam} {et~al.}(2006){Gal-Yam}, {Ofek}, {Poznanski}, {Levinson},
  {Waxman}, {Frail}, {Soderberg}, {Nakar}, {Li}, \&
  {Filippenko}}]{2006ApJ...639..331G}
{Gal-Yam}, A., {et~al.} 2006, \apj, 639, 331

\bibitem[{{Gezari} {et~al.}(2006){Gezari}, {Martin}, {Milliard}, {Basa},
  {Halpern}, {Forster}, {Friedman}, {Morrissey}, {Neff}, {Schiminovich},
  {Seibert}, {Small}, \& {Wyder}}]{2006ApJ...653L..25G}
{Gezari}, S., {et~al.} 2006, \apjl, 653, L25

\bibitem[{{Gezari} {et~al.}(2009){Gezari}, {Heckman}, {Cenko}, {Eracleous},
  {Forster}, {Gon{\c c}alves}, {Martin}, {Morrissey}, {Neff}, {Seibert},
  {Schiminovich}, \& {Wyder}}]{2009ApJ...698.1367G}
---. 2009, \apj, 698, 1367

\bibitem[{{Giannios} \& {Metzger}(2011)}]{2011arXiv1102.1429G}
{Giannios}, D., \& {Metzger}, B.~D. 2011, ArXiv e-prints

\bibitem[{{Ivison} {et~al.}(2007){Ivison}, {Chapman}, {Faber}, {Smail},
  {Biggs}, {Conselice}, {Wilson}, {Salim}, {Huang}, \&
  {Willner}}]{2007ApJ...660L..77I}
{Ivison}, R.~J., {et~al.} 2007, \apjl, 660, L77

\bibitem[{{Levinson} {et~al.}(2002){Levinson}, {Ofek}, {Waxman}, \&
  {Gal-Yam}}]{2002ApJ...576..923L}
{Levinson}, A., {Ofek}, E.~O., {Waxman}, E., \& {Gal-Yam}, A. 2002, \apj, 576,
  923

\bibitem[{{Rees}(1988)}]{1988Natur.333..523R}
{Rees}, M.~J. 1988, \nat, 333, 523

\bibitem[{{Strubbe} \& {Quataert}(2010)}]{2010arXiv1008.4131S}
{Strubbe}, L.~E., \& {Quataert}, E. 2010, ArXiv e-prints

\bibitem[{{van Velzen} {et~al.}(2010){van Velzen}, {Farrar}, {Gezari},
  {Morrell}, {Zaritsky}, {Ostman}, {Smith}, \& {Gelfand}}]{2010arXiv1009.1627V}
{van Velzen}, S., {Farrar}, G.~R., {Gezari}, S., {Morrell}, N., {Zaritsky}, D.,
  {Ostman}, L., {Smith}, M., \& {Gelfand}, J. 2010, ArXiv e-prints

\bibitem[{{Wang} \& {Merritt}(2004)}]{2004ApJ...600..149W}
{Wang}, J., \& {Merritt}, D. 2004, \apj, 600, 149

\bibitem[{{Welch} {et~al.}(2009){Welch}, {Backer}, {Blitz}, {Bock}, {Bower},
  {Cheng}, {Croft}, {Dexter}, {Engargiola}, {Fields}, {Forster},
  {Gutierrez-Kraybill}, {Heiles}, {Helfer}, {Jorgensen}, {Keating}, {Lugten},
  {MacMahon}, {Milgrome}, {Thornton}, {Urry}, {van Leeuwen}, {Werthimer},
  {Williams}, {Wright}, {Tarter}, {Ackermann}, {Atkinson}, {Backus}, {Barott},
  {Bradford}, {Davis}, {Deboer}, {Dreher}, {Harp}, {Jordan}, {Kilsdonk},
  {Pierson}, {Randall}, {Ross}, {Shostak}, {Fleming}, {Cork}, {Vitouchkine},
  {Wadefalk}, \& {Weinreb}}]{2009IEEEP..97.1438W}
{Welch}, J., {et~al.} 2009, IEEE Proceedings, 97, 1438

\end{thebibliography}

\begin{deluxetable}{lrrrrrrr}
\tablecaption{Surveys with Long Timescale Sensitivity\label{tab:surveys}}
\tablehead{ \colhead{Name} & \colhead{$f$} & \colhead{$F_{lim}$}
& \colhead{$\Omega$} & \colhead{$N_{e}$} & \colhead{$N_{can}$} & \colhead{$D_{lim}$} & \colhead{$r_{TD}$} \\
                           & \colhead{ (GHz) } & \colhead{ (mJy) } & \colhead{(sr)} & &  & \colhead{(Gpc)} & \colhead{($10^{-7} {\rm\ y^{-1}\ Mpc^{-3}}$)}
}
\startdata
VLA              &   5.0 &    0.1 &  3.9e-06 &    20 &    0 & 2.99 &  $ <    29$  \\ 
3C286            &   1.4 &   70.0 &  1.6e-04 &    23 &    0 & 0.08 &  $ < 29042$  \\ 
PiGSS-I          &   3.1 &    2.0 &  3.0e-03 &     1 &    0 & 0.59 &  $ <    96$  \\ 
ATATS-I          &   1.4 &  230.0 &  2.1e-01 &     1 &    0 & 0.05 &  $ <  3113$  \\ 
MOST             &   0.8 &   14.0 &  5.5e-01 &     1 &    2 & 0.16 &  $ <    17$  \\ 
FIRST-NVSS       &   1.4 &    6.0 &  1.9e-01 &     1 &    0 & 0.28 &  $ <    14$  \\ 
\enddata
\end{deluxetable}

\end{document}